\documentclass[a4paper]{elsarticle}

\usepackage{amsmath,amssymb,amsthm,mathtools}
\usepackage{graphicx}
\usepackage{verbatim}
\usepackage{ifthen}
\numberwithin{equation}{section}

\biboptions{compress}

\makeatletter
\def\ps@pprintTitle{%
  \let\@oddhead\@empty
  \let\@evenhead\@empty
  \def\@oddfoot{\reset@font\hfil\thepage\hfil}
  \let\@evenfoot\@oddfoot
}
\makeatother

\newtheorem{theorem}{Theorem}[section]

\newcommand{\Z}{\mathbb{Z}}
\newcommand{\R}{\mathbb{R}}
\newcommand{\C}{\mathbb{C}}
\newcommand{\F}{\mathbb{F}}

\newcommand{\br}[1]{\left(#1\right)}
\newcommand{\pd}[2]{ \frac{\partial #1}{\partial #2} }

\newcommand{\pdd}[3]{
\ifthenelse{\equal{#3}{}}{\frac{\partial^2 #1}{\partial {#2}^2}}{\frac{\partial^2 #1}{\partial #2 \partial #3}}
}

\newcommand{\bmat}[1]{\begin{bmatrix} #1 \end{bmatrix}}
\newcommand{\pmat}[1]{\begin{pmatrix} #1 \end{pmatrix}}

\newcommand{\pdddd}[5]{
\ifthenelse{\equal{#5}{} \and \equal{#4}{} \and \equal{#3}{}}{\frac{\partial^3 #1}{\partial #2 \partial #3 \partial #4 \partial #5}}{\frac{\partial^4 #1}{\partial #2 \partial #3 \partial #4 \partial #5}}
}

\newcommand{\m}{m}
\newcommand{\n}{n}

\newcommand{\prm}{{\pi(\m)}}
\newcommand{\efm}{\m^*}
\newcommand{\dfm}{\m_*}
\newcommand{\prn}{{\pi(\n)}}
\newcommand{\efn}{\n^*}
\newcommand{\dfn}{\n_*}
\newcommand{\pro}{{\pi(o)}}

\newcommand{\charspaces}[2]{\mathcal{S}_{#1}^{(#2)}}
\newcommand{\txi}{\tilde{\Xi}}

\DeclareMathOperator{\diag}{diag}
\DeclareMathOperator{\mspan}{span}
\DeclareMathOperator{\linind}{lin. ind.}
\DeclareMathOperator{\sgrp}{Sp}

\journal{Nuclear Physics B}

\begin{document}

\begin{frontmatter}

\author[kdvi,itep]{Petr Dunin-Barkowski}
\ead{P.Dunin-Barkovskiy@uva.nl, barkovs@itep.ru}
\address[kdvi]{Korteweg-de~Vries Institute for Mathematics, University of Amsterdam, the Netherlands}
\address[itep]{Institute for Theoretical and Experimental Physics, Moscow, Russia}
\author[kdvi,itep]{Alexey Sleptsov}
\ead{sleptsov@itep.ru}
\author[kdvi]{Abel Stern\corref{cor1}}
\ead{abel.stern@gmail.com}
\title{NSR superstring measures in genus 5}
\cortext[cor1]{Corresponding author. Postal address of Korteweg-de Vries Institute: P.O. Box 94248, 1090 GE Amsterdam, The Netherlands. Phone number: +31 20-525 5217 (secretary) }

\date{August 10th, 2012}

\begin{abstract}
Currently there are two proposed ans\"atze for NSR superstring measures: the Grushevsky ansatz and the OPSMY ansatz, which for genera $g\leq 4$ are known to coincide. However, neither the Grushevsky nor the OPSMY ansatz leads to a vanishing two point function in genus four, which can be constructed from the genus five expressions for the respective ans\"atze. This is inconsistent with the known properties of superstring amplitudes.

In the present paper we show that the Grushevsky and OPSMY ans\"atze do not coincide in genus five. Then, by combining these ans\"atze, we propose a new ansatz for genus five, which now leads to a vanishing two-point function in genus four. We also show that one cannot construct an ansatz from the currently known forms in genus 6 that satisfies all known requirements for superstring measures.
\end{abstract}


\end{frontmatter}

\section{Introduction} \label{sec:introduction}
In perturbative superstring theory in the NSR formalism, scattering amplitudes can be represented
as integrals over the moduli space of super Riemann surfaces $\mathfrak{M}_g$
with respect to a certain measure. Therefore, this \emph{superstring measure} is one of its main ingredients. 

For the genus 0 and 1 cases it was known from the start \cite{green1987superstring,green1988superstring} that the measure can be 
written as a collection of modular forms, for different subgroups of the modular group, on the moduli space of ordinary Riemann surfaces. In a prominent series of papers \cite{dhokerphong1,dhokerphong2,dhokerphong3,dhokerphong4,dhokerphong5,dhokerphong6,dhokerphong7,dhokerphong8,dhokerphong9} E. D'Hoker and D. Phong showed that this is true for genus 2 as well, and moreover they obtained explicit expressions for these measures in terms of theta constants. 

 One thus hopes (for the history, cf. \cite{morozovrevis}) that one can (by integrating out the odd moduli) move to a measure on the moduli space of ordinary Riemann surfaces $\mathcal{M}_g$ in \emph{all} genera. This would be very useful as actual calculations on the moduli space of \emph{super} Riemann would be much more complicated.

Finding a way to integrate out the odd moduli has proven to be exceedingly difficult already in genus 2, as can be seen from the fact that it took D'Hoker and Phong twenty years to succeed in doing so. Therefore, an alternative approach was proposed \cite{cdg1,cdg2,belavinknizhnik1,belavinknizhnik2,knizhnik1,belavinknizhnikmorozovperelomov,moore,alvarezgaume,morozovexplicit,morozovanalytical,dhokerphong10,dhokerphong11} where instead of explicit calculation, ans\"{a}tze were made based on supposed requirements for the measure.

If the superstring measure can be written as a measure on $\mathcal{M}_g$, the formula for the superstring partition function at the $g$-loop level will be as follows \cite{dhokerphong4}:
\begin{align}
Z_g &= \int_{\mathcal{J}_g / \sgrp(2g,\Z)} (\det \Im( \tau^{(g)}))^{-5} d\mu(\tau) \wedge \overline{d\mu(\tau)} \label{eq:partfunc} \\
d\mu(\tau) &= \sum_m d\mu[m] (\tau) \nonumber
\end{align}
where summation is over even spin structures $m$ on the Riemann surface, which are the same as even theta characteristics \cite{atiyah1971riemann}. The factor of $(\det \Im( \tau))^{-5}$ results from integrating over internal momenta, with a power of half the critical dimension, as in the case for the bosonic string \cite{dhokerphong12}. The $d\mu[\m]$ are measures on the Jacobian locus $\mathcal{J}_g$, the subset of all period matrices $\tau$ inside the Siegel half-space. They are labelled by theta characteristics $m \in \F_{2}^{(2g)}$.  In order for the right hand side of \eqref{eq:partfunc} to be a well-defined integral over $\mathcal{J}_g / \sgrp(2g,\Z)$, the full measure (being defined on $\mathcal{J}_g$) has to be invariant under the action of the modular group $\sgrp(2g,\Z)$. Since $\det \Im(\tau)$ transforms as a modular form of weight $-2$, we see that all $d\mu[\m]$ must transform as modular forms of weight $-5$ with respect to the subgroups $\Gamma[m]$ conjugate to $\Gamma(1,2) \subset \sgrp(2g,\Z)$, the subgroup that fixes the zero theta characteristic (see section \ref{sec:definitions}).

It has been conjectured (see \cite{cdg1} for a discussion) that the NSR measures $d \mu[\m]$ can be written as a product of the Mumford measure for the critical bosonic string $d \mu$ (which is of weight $-13$) and for each characteristic a modular form $\Xi[\m]$ of weight 8 on the Siegel upper half-space:
\begin{align}
 d \mu[\m] = \Xi[\m] d\mu.
\end{align}
The conditions to which the measure, if the above conjecture holds, must conform are the following:
\begin{enumerate}[a)]
 \item The forms $\Xi[\m]$ are modular forms of weight 8 with respect to $\Gamma[m]$ when restricted to the Jacobian locus (the closure of the subspace of period matrices inside the Siegel upper half-space).
 \item The forms satisfy the factorization (splitting) property on block-diagonal period matrices: $\Xi_{\m \times \n}^{(g)}\pmat{ \tau^{(g-k)} & 0 \\ 0 & \tau^{(k)} } = \Xi_{\m}^{(g-k)}\br{\tau^{(g-k)}} \Xi_{\n}^{(k)}\br{\tau^{(k)}}$.
 \item The trace (the cosmological constant) should vanish, i.e. $\sum_\m \Xi[\m] = 0$. Also, the trace of the $1,\ldots,3$-point functions $\sum_\m A_k[\m]$ should vanish\footnote{Naturally, this can only yield a condition on $\Xi[m]$ when we know how the 2- and 3-point functions can be obtained from the measure. However, Matone and Volpato recently proposed how to do this in some
cases; see \cite{matone2010getting} for the results on two-point
functions. In \cite{matone2009superstring} they show that the connected part of the 3-point function for the Grushevsky ansatz in genus 3 does not vanish, and argue that it is cancelled by the disconnected part.}, cf. \cite{martinec1,martinec2}.
 \item In genus 1 the ansatz should conform to the known answer.
\end{enumerate}

In genus $g \leq 3$ it is known \cite{dalla2009siegel} that there is a unique way of satisfying these constraints , so the conjecture holds, but in general for higher genera it is not known a priori whether a suitable modular form on the Siegel half-space exists. The ratio of $d \mu[\m]$ to $d \mu$ may very well only be holomorphic on the Jacobian locus and be meromorphic elsewhere. The Jacobian locus has positive codimension from genus 4 on. As the dimension of the space of modular forms on the Jacobian locus with respect to the relevant groups is not known, it is unclear whether the above conditions will lead to a unique definition of the forms $\Xi[\m]$. In the present paper we show that combinations of the known modular forms are not suitable for satisfying all the above conditions in higher genera.

Two sets of ans\"{a}tze were proposed. First, an ansatz was proposed for genus 3 by S.L. Cacciatori, F. Dalla Piazza and B. van Geemen in \cite{cdg1}. It was then elegantly generalized to genera 4 and above (subject to certain forms being well-defined) by S. Grushevsky in \cite{grushevsky}. It was then shown by Salvati Manni that the Grushevsky ansatz is well-defined in genus 5 \cite{salvati2008remarks}, and Salvati Manni and Grushevsky modified the original ansatz to obtain a vanishing cosmological constant in genus 5 \cite{grushevsky2011superstring}. However, in genus 6 there is yet no reason to believe that the ansatz is well-defined and the modification in genus 5 spoils the genus 6 factorization property. Then, the second ansatz was formulated in terms of theta series for 16-dimensional self-dual lattices by M. Oura, C. Poor,R. Salvati Manni and D. Yuen (OPSMY) in \cite{oura2010modular}. This second ansatz, however, is only defined for genera $g \leq 5$.

Both ans\"{a}tze do, in their final forms, satisfy requirements a),b) and d), and have vanishing cosmological constant in genera $1,\ldots,5$. However, it was shown by M. Matone and R. Volpato in \cite{matone2010getting} that the genus 4 two-point function obtained by degeneration from the OPSMY ansatz in genus 5 does not vanish, contrary to requirement c). The results of the present paper imply that the same problem occurs with the Grushevsky ansatz as well.

The paper \cite{latticevsriemann} compares the modular forms $G_p^{(g)}$ and $\vartheta_p^{(g)}$, from which the Grushevsky and OPSMY ans\"{a}tze were constructed. $G_p^{(g)}$ are certain polynomials in fractional powers of theta constants, whilst $\vartheta_p^{(g)}$ are genus $g$ theta series of 16-dimensional unimodular lattices, see section \ref{sec:definitions} for definitions. For all but one $p$ (where $0 \leq p \leq 7$) it was shown that $\vartheta_p^{(g)}$ was expressible as a linear combination of the $G_i^{(g)}$, for \emph{all} genera. This implies that both ans\"{a}tze are identical up to and including genus 4. For genus 5 and above, however, the question whether $G_5^{(g)}$ and $\vartheta_5^{(g)}$ agree on the Jacobian locus remained open.

In summary, there are two ans\"{a}tze, defined for genera $g \leq 5$, which were shown to be identical for $g \leq 4$, although it was unknown until the present paper whether they agree in genus 5, and both ans\"{a}tze suffer from the same problem of a not identically vanishing two-point function in genus 4.

A natural question, then, became whether these ans\"{a}tze do in fact coincide for genus $g =5 $ and if not, what can be done by combining their building blocks.

\paragraph{Results}
In the present paper (at the end of section \ref{sec:degeneration}) we show that in fact, for genus $g \geq 5$, on the Jacobian locus, $G_g^{(g)}$ and $\vartheta_5^{(g)}$ do \emph{not} agree. This implies that the OPSMY and Grushevsky ans\"{a}tze differ in genus 5. We use the fact that $\vartheta_5^{(5)} - G_5^{(5)}$ is not identically zero on the Jacobian locus to present a modified genus 5 ansatz,
\begin{align}
 \txi &:= \Xi^{(5)}_{OPSMY} - \frac{222647008}{217} \br{\vartheta_6^{(5)} - \vartheta_7^{(5)}} + \frac{77245568}{17} \br{\vartheta_5^{(5)} - G_5^{(5)} }.
\end{align}
We prove the vanishing of both the genus 5 cosmological constant and the genus 4 two-point function, obtained from degeneration, for this modified ansatz. The second statement holds assuming that the Matone-Volpato method \cite{matone2010getting} is the correct way to obtain the genus 4 two-point function from the genus 5 zero-point function. Then, we look at the situation in genus 6. We show that it is not possible to construct a genus 6 ansatz from the currently known forms that satisfies all properties. To be precise, condition c) cannot be satisfied.

\paragraph{Remark} Because $\vartheta_5^{(g)} - G_g^{(g)}$ is known to factorize to the genus 4 Schottky form, it cannot vanish identically on the Jacobian locus for all $g$, as it would then be a stable Schottky form, which was shown to be impossible recently \cite{codogni2011schottky}.

\paragraph{Structure of the present paper} The paper is organized as follows: in section \ref{sec:definitions} we define the modular forms used in the OPSMY and Grushevsky ans\"{a}tze and list the known relations between those sets of forms. In section \ref{sec:degeneration} we expand $\vartheta_5^{(5)} - G_5^{(5)}$ in a perturbative series by contracting one handle of the curves and show that this series does not vanish on the entire Jacobian locus, which means $\vartheta_5^{(5)}-G_5^{(5)}$ is not identically zero there. In section \ref{sec:trace} we calculate the trace (the summation $\sum_\m f[\m]$ over even characteristics) of this function. We need this to prove that the cosmological constant for our modified ansatz in genus 5 vanishes. In section \ref{sec:difference} we compare $\vartheta_5^{(5)} - G_5^{(5)}$ with other modular forms to show it is not equal to one of the already known forms. In section \ref{sec:twopointfunction} we look at the two-point function in genus 4 obtained by degenerating the genus
5 ansatz $\Xi^{(5)}_{OPSMY} + c\br{\vartheta_6^{(5)} - \vartheta_7^{(5)}} + d \br{\vartheta_5^{(5)}
 - G_5^{(5)}}$, by the method used in \cite{matone2010getting}. We show that this, together with the condition of vanishing genus 5 cosmological constant leads to our main formula \eqref{eq:mainformula}: a unique ansatz built from the known modular forms in
genus 5.  In section \ref{sec:genus6} we discuss the factorization property for any genus 6 ansatz implied by our proposed modification for genus 5. We show that it cannot be satisfied using only the known forms. Finally, in section \ref{sec:discussion} we briefly discuss our results.

\section{Definitions: the modular forms from OPSMY and Grushevsky} \label{sec:definitions}
The superstring ans\"{a}tze are linear combinations of modular forms of weight 8 on the Jacobian locus. Here, we will define the relevant concepts.

Let $\mathcal{H}_g$ be the Siegel upper half-space, i.e. the set of complex symmetric $g \times g$-matrices for which the imaginary part is positive definite. Let $\sgrp(2g,\Z)$ be the symplectic group of degree $2g$ over $\Z$, here called the \emph{modular group} $\Gamma_g$. The modular group acts on the Siegel upper half-space through modular transformations, defined as follows: let $\gamma = \pmat{A & B \\ C & D} \in \Gamma_g$. Then set
\begin{align}
 \gamma \circ \tau := (A \tau + B)(C \tau + D)^{-1}, \qquad \tau \in \mathcal{H}_g
\end{align}
Hence we can also define an action on functions on the Siegel upper half-space. The action is defined as follows, for a given $k$:
\begin{align}
 (f|_k \gamma)(\tau) := \det(C\tau + D)^{-k} f\br{\gamma \circ \tau}.
\end{align}

Theta characteristics are elements of $\F_2^{(2g)}$ which we will write as $\m$ or as $\bmat{\epsilon \\ \delta}$, where $\epsilon,\delta \in \F_2^g$; see the introduction. In the literature these are sometimes called semi-integer characteristics to distinguish them from the rational characteristics appearing elsewhere. We will view $\F_2$ as a $\C$-module with the obvious product $\bar{1} \cdot z = z$, $\bar{0} \cdot z = 0$.

The theta characteristics are called even (resp. odd) if the standard inner product $\epsilon \cdot \delta$ is even (resp. odd).

For a theta characteristic $\m = \bmat{\epsilon \\ \delta} $, $\epsilon = (\epsilon_1,\ldots,\epsilon_g)$, $\delta = (\delta_1, \ldots, \delta_g)$ we will denote $\efm = \epsilon_1$, $\dfm = \delta_1$, as these components will frequently pop up in the Fourier-Jacobi expansion.

The modular group also acts on the theta characteristics, as follows: for $\gamma$ as above, let (with ordinary matrix multiplication and addition in $\F_2$)
\begin{align}
\gamma [\m] := \pmat{ D & -C \\ -B & A} \bmat{\epsilon \\ \delta} + \bmat{ \diag(CD^T) \\ \diag(AB^T) }.
\end{align}
Let $\Gamma(1,2)_g$ be the subgroup of $\Gamma_g$ that fixes the zero characteristic by the above action. Then, we can mark each subgroup conjugate to $\Gamma(1,2)_g$ with a theta characteristic $\m$ by the action of the conjugating element on the zero characteristic, that is, we will write $\Gamma[\m]_g = \gamma \Gamma(1,2)_g \gamma^{-1}$ iff $\gamma[0] = \m$.

A holomorphic function $f$ on the Siegel upper half-space $\mathcal{H}_g$ is called a modular form of weight $k$ with respect to a certain subgroup $G \subset \Gamma_g$ if the following holds:
\begin{align}
 \forall \gamma \in G,\quad (f|_k \gamma) = f.
\end{align}

Let $C$ be a Riemann surface of genus $g$. Let us pick a basis for the homology group $H_1(C,\Z)$. Then we have the period matrix $\tau \in \mathcal{H}_g$ of $C$. Thus we have a map $\tau: \mathcal{M}_g \rightarrow \mathcal{H}_g / \Gamma_g$, where $\mathcal{M}_g$ is the moduli space of Riemann surfaces of genus $g$. The subset $\mathcal{J}_g \subset \mathcal{H}_g$ of all possible period matrices is called the Jacobian locus and $\mathcal{J}_g \subsetneq \mathcal{H}_g$ for $g \geq 4$. We will write $\omega_i$ for the $i$th holomorphic differential in the basis corresponding to the period matrix. Also, we use the Abel-Jacobi map $A$, constructed from the same basis mentioned above, and we will write $A_{pq} := A(p) - A(q)$. For details, we refer to \cite{griffithsharris}.

The OPSMY ansatz from \cite{oura2010modular} is constructed using lattice theta series, defined as follows for any lattice $\Lambda \subset \R^n$:
\begin{align}
\vartheta^{(g)}_{\Lambda}(\tau) := \sum_{p_1,\ldots,p_g \in \Lambda} e^{\pi i \sum_{i,j} \tau_{ij} p_i \cdot p_j}
\end{align}
The theta series of self-dual $8n$-dimensional lattices provide us with modular forms with respect to $\Gamma(1,2)_g$ of weight $4n$ , which are in addition modular with respect to the entire group $\Gamma_g$ if the lattice is even.

There are 8 self-dual lattices of dimension 16 \cite{conway1999sphere}. We will introduce shorthand notation for the corresponding theta series, in line with \cite{latticevsriemann},
\begin{center}
\begin{tabular}{|c|c|c|}
\hline
Notation&Lattice&Gluing vectors\\
\hline
$\vartheta_0$&$\Z^{16}$&-\\
\hline
$\vartheta_1$&$\Z^8\oplus E_{8}$&-\\
\hline
$\vartheta_2$&$\Z^4\oplus D_{12}^+$&$(0^4,\frac12^{12})$\\
\hline
$\vartheta_3$&$\Z^2\oplus \br{E_7\oplus E_7}^+$&$(\frac14^6,-\frac34^2,\frac14^6,-\frac34^2)$\\
\hline
$\vartheta_4$&$\Z\oplus A_{15}^+$&$(\frac14^{12},-\frac34^4),(\frac12^8,-\frac12^8),(\frac34^4,-\frac14^{12})$
\\
\hline
$\vartheta_5$&$\br{D_8\oplus D_8}^+$&$(\frac12^8,0^7,1)$
\\
\hline
$\vartheta_6$&$E_8 \oplus E_8$&-
\\
\hline
$\vartheta_7$&$D_{16}^+$&$(\frac12^{16})$ \\
\hline
\end{tabular}
\end{center}

The lattices denoted by $\Lambda^+$ are obtained by taking the union $\Lambda \cup (v_1 + \Lambda) \cup \ldots$ of the lattice with itself, shifted by the gluing vectors $v_i$. For example, $E_8 = D_8^+ = D_8 \cup ((\frac12^8) + D_8)$.

The Grushevsky ansatz, from \cite{grushevsky}, is instead built using Riemann theta functions, defined as follows for a theta characteristic $\m = \bmat{\epsilon \\ \delta}$, here regarded as a vector in $\C^{2g}$,
\begin{align}
 \theta \bmat{\epsilon \\ \delta} (z,\tau) := \sum_{n \in \Z^g} \exp \left\{ \pi i \left( n + \frac12 \epsilon \right)^t \tau \left( n + \frac12 \epsilon\right) +  2 \pi i \left( n + \frac12 \epsilon \right)^t \left(z + \frac12 \delta \right) \right\}.
\end{align}
Riemann theta functions for $z=0$ are called Riemann theta constants. The Riemann theta constants of odd characteristics are zero identically on $\mathcal{H}_g$. We will write $\theta_\m := \theta\bmat{\epsilon \\ \delta}(0,\tau)$.

The modular forms used in \cite{grushevsky} are defined as follows. Let $V \subset \F_2^{(2g)}$ be a set of characteristics in genus $g$. Then, we define
\begin{align}
 P (V) := \prod_{\m \in V} \theta_\m\,.
\end{align}
Now, define $\charspaces{p}{g}$ to be the set of all $p$-dimensional linear subspaces of $\F_2^{(2g)}$.  Then, we define the Grushevsky forms $\{G_p^{(g)},0\leq p\leq g\in \Z \}$ as follows:
\begin{align}
  G_p^{(g)} := \sum_{V \in \charspaces{p}{g}} P(V)^{2^{4-p}}.
\end{align}
These forms are modular with respect to $\Gamma(1,2)_g$ and of weight 8.
Note that this normalization differs from that in \cite{latticevsriemann} by a factor of $\br{ 2^{\frac{p(p-1)}2} \prod_{i=1}^p \br{2^i-1}}$, taken to be 1 for $p=0$:
\begin{align}
G_p^{(g)} = \br{ 2^{\frac{p(p-1)}2} \prod_{i=1}^p \br{2^i-1}} \sum_{\substack{e_1,\ldots,e_p \in \F_2^{(2g)} \\ e_1,\ldots,e_p \linind}}  P(\mspan\{e_1,\ldots,e_p\})^{2^{4-p}}
\end{align}

From \cite{latticevsriemann} we have several linear dependencies between lattice theta series and Riemann theta constants. In the present notation they look as follows, for $p \leq 4$:
\begin{align}
 G_p^{(g)} &= \sum_{k=0}^p (-1)^{k+p} \cdot 2^{\frac{k(k +2(g-p) + 1)}2} \cdot \br{ \prod_{i=1}^k (2^i - 1) \prod_{i=1}^{p-k} (2^i -1) }^{-1} \vartheta_k^{(g)}
\end{align}
where $\prod_{i=1}^k (2^i - 1)$ is taken to be 1 for $k=0$.

Throughout the paper we will denote
\begin{align}
      f^{(g)} := \vartheta_5^{(g)} - G_g^{(g)} \\
      J^{(g)} := \vartheta_6^{(g)} - \vartheta_7^{(g)}.
      \end{align}
It was shown in \cite{latticevsriemann} that $f^{(g)}$ vanishes identically on the Jacobian locus $\mathcal{J}_g$ for $g \leq 4$. In the present paper we show that $f^{(5)}$ does not vanish identically on $\mathcal{J}_5$.

\section{Degeneration} \label{sec:degeneration}

The conjecture which we investigate and disprove in this section is whether $G_5^{(5)}$ and $\vartheta_5^{(5)}$ agree on the Jacobian locus~$\mathcal{J}_5$.

To show that $f^{(5)}= \vartheta_5^{(5)} - G_5^{(5)}$ is not identically vanishing on $\mathcal{J}_5$, we use the procedure used by Grushevsky and Salvati Manni in \cite{grushevsky2011superstring}, which is based on a theorem by Fay \cite{fay}. Our motivation for using this method is that in \cite{grushevsky2011superstring} it was succesfully applied to show that $J^{(5)}$ does not vanish everywhere on $\mathcal{J}_5$.

The method is as follows: we will take a 1-parameter family of Riemann surfaces $C_s \subset \mathcal{M}_5$, with parameter $s$, which, as $s \rightarrow 0$, degenerates to a genus 4 surface $C$ with two nodes $p$ and $p'$, inside the boundary divisor $\delta_0 \subset \overline{\mathcal{M}_5}$. We take a Taylor series in $s$ as $s \rightarrow 0$ of $f^{(5)}$ and show that the first-order term in $s$ is not identically vanishing. Since $f^{(5)}$ is holomorphic on $\mathcal{J}_5$, this implies that $G_5^{(5)}$ and $\vartheta_5^{(5)}$ are not identically equal on $\mathcal{J}_5$.

As shown in \cite{fay} we can take such a family of surfaces that their period matrices $\tau_s$ have the following form:
\begin{align}
 \tau_s =  \pmat{ \lambda & z \\ z^t & \tau }=  \pmat{\ln s + c_1 + c_2 s & A^t_{pp'} + \frac14 s (\omega(p) - \omega(p'))^t \\ A_{pp'} + \frac14 s (\omega(p) - \omega(p'))& \tau_0 + s\,\sigma }
\end{align}
for some constants $c_1$ and $c_2$, where $\tau_0$ is the period matrix of $C_0$ and
\begin{align*}
  \sigma_{ij} &:= \frac14 \br{\omega_i(p) - \omega_i(p')}\br{\omega_j(p) - \omega_j(p')}, \qquad i,j \leq 4.
\end{align*}
Define as elsewhere in the literature, for legibility,
\begin{align}
q := e^{2 \pi i \lambda}.
\end{align}
Now, if we obtain the Fourier-Jacobi expansions of $G_5^{(5)}$ and $\vartheta_5^{(5)}$, we can use this to express the forms evaluated in $\tau_s$ as series in $s$. That is, for any function $f$ on $\mathcal{J}_5$ that is holomorphic on a neighbourhood of the curve $\{\tau_s\} \subset \mathcal{J}_5$, if
\begin{align}
 f(\tau_s) = f_0(\tau) + q f_1(\tau,z) + O(q^2) \label{eq:fjexpansion}
\end{align}
we have
\begin{align}
 f(\tau_s) = f_0(\tau_0) + s \br{ \sum_{i \leq j}^4 \pd{f_0(\tau)}{\tau_{ij}} \sigma_{ij}(p,p') + f_1(\tau,z)} + O(s^2). \label{eq:sseries}
\end{align}

We will express the first terms above in a Taylor series. We take for a local chart $x$ the parameter $u = x(p) - x(p')$ near $u=0$ and calculate, following \cite{grushevsky2011superstring},
 \begin{align}
 \sigma_{ij}(p,p') &= S_{ij} +  O(u^4) \\
 S_{ij} :&= u^2 \frac14 \pd{\omega_i(p)}{x} \pd{\omega_j(p)}{x} + u^3 \frac12  \pdd{\omega_i(p)}{x}{} \pd{\omega_j(p)}{x}
\end{align}
and therefore, if $\pd{f_1}{z_i}$ and $\frac{\partial f_1^3}{\partial z_i \partial z_j \partial z_k}$ vanish,
\begin{align}
 f(\tau_s(p,p')) = f_0 (\tau_0) + s \sum_{i \leq j}^4 \br{ u^2 \pdd{f_1}{z_i}{z_j} \omega_i(p) \omega_j(p) + \pd{f_0}{\tau_{ij}} S_{ij} +  O(u^4)} + O(s^2). \label{eq:useries}
\end{align}
These series for $G_5^{(5)}$ and $\vartheta_5^{(5)}$, then, can finally be shown to disagree, by an argument used in \cite{grushevsky2011superstring}.

\subsection{The expansion of $G_5^{(5)}$}
To determine the degeneration of $G_5^{(5)}$ and $\vartheta_5^{(5)}$ we will here take the Fourier-Jacobi expansion \eqref{eq:fjexpansion} of $G_5^{(5)}$. That is, we will express $G_5^{(5)}(\tau_s)$ in the limit $\lambda \rightarrow \infty$. Also, we will calculate $\pdd{G_{5,1}^{(5)}}{z_i}{z_j}$ where $G_{5,1}^{(5)}$ stands for the $q$-linear term in the Fourier-Jacobi expansion of $G_5^{(5)}$.

\subsubsection{Expanding $P(V)^{\frac12}$}
First, we will calculate the Fourier-Jacobi expansion of the summands $P(V)^{\frac12}$ for $V \in \charspaces{5}{5}$. Recall that for $\m = \bmat{\epsilon \\ \delta} $, $\epsilon = (\epsilon_1,\ldots,\epsilon_g)$, $\delta = (\delta_1, \ldots, \delta_g)$ we will denote $\efm = \epsilon_1$, $\dfm = \delta_1$. Let $\pi$ be the projection from $\F_2^{(2g)}$ to  $\F_2^{(2g-2)}$ by sending $\m = \bmat{\epsilon_1 &  \epsilon_2 & \ldots & \epsilon_g \\ \delta_1 & \delta_2 & \ldots & \delta_g}$ to $\pi(\m) = \bmat{\epsilon_2 & \ldots & \epsilon_g \\ \delta_2 &  \ldots &  \delta_g} \in \F_2^{(2g-2)}$. We will use the known formulae for the Fourier-Jacobi expansion of theta constants, which look as follows \cite{van1984siegel}:
\begin{align}
\theta \bmat{ 0 & \epsilon \\ * & \delta} \pmat{ \lambda & z^t \\ z & \tau } &= \theta \bmat{\epsilon \\ \delta}
+ 2\, q^{1/2} e^{\pi i *}  \, \theta \bmat{\epsilon \\ \delta} \left(z, \tau \right)  + O(q^2) \label{eq:FJdelta0}
 \\
\theta \bmat{ 1 & \epsilon \\ * & \delta}  \pmat{ \lambda & z^t \\ z & \tau} &= 2 \, q^{1/8}\, e^{\pi i * /2}\,  \theta \bmat{\epsilon \\ \delta}(\frac{z}2, \tau) + O(q^{9/8})  \label{eq:FJdelta1}.
\end{align}

As each component of the characteristics contained in $V$ can be either 0 or 1, and $P(V)^{\frac12}$ vanishes if $V$ contains any odd characteristics, we can distinguish three kinds of subspaces $V$ having different expansions of $P(V)^{\frac12}$. For each of these we will calculate $P(V)^{\frac12}$ and $\pdd{}{z_i}{z_j} P(V)^{\frac12}$ to first order in $q$.

\begin{enumerate}
 \item First, we consider subspaces containing only characteristics of the form $\m =\bmat{ 0 & \epsilon \\ * & \delta} $. Thus, expanding $P(V_1)$ for $V_1$ of this type, using \eqref{eq:FJdelta0}, we get
  \begin{align}
  P(V_1) &= \prod_{\m \in V_1} \theta_\prm + 2 q^{1/2}
	\sum_{\m \in V_1}
	e^{\pi i \dfm} \theta_\prm (\tau,z)
\prod_{\substack{\n \in V_1 \\ v \neq e}} \theta_\prn \nonumber \\ & \qquad + 2  q
\sum_{\substack{ \m,\n \in V_1 \\ \m \neq \n } }
		e^{\pi i \br{\dfm + \dfn }}\theta_\prm (\tau,z) \theta_\prn (\tau,z)
\prod_{\substack{o \in V_1 \\ o \neq \m \\ o \neq \n }} \theta_\pro + O(q^2). \label{eq:firstkindpv}
\end{align}

 For such $V_1$, the image $\pi(V_1)$ is totally isotropic, and therefore the space $\pi(V_1)$ has maximal dimension 4.  Because additionally the kernel of $\pi$ has a maximal dimension of 1 (only $\dfm$ can be picked freely), the $\prm$ are necessarily pairwise equal, the corresponding pairs of $\m$ differing only in their $\dfm$. Define $\Delta \in \mathcal{F}_g^{(2g)} = \bmat{0 & 0^{g-1} \\ 1 & 0^{g-1} }$. 
 The above consideration shows that $\m + \Delta$ is contained in $V_1$. Unless $\m = \n + \Delta$, each term $\m,\n$ in the summation in the third term from \eqref{eq:firstkindpv} will be canceled by a $\m+\Delta,\n$ term. Combining these facts, we can rewrite the above formula as follows:
 \begin{align}
  P(V_1) = \prod_{\m \in \pi(V_1)} \theta_\m^2 -4 q \sum_{\m \in \pi(V_1)} \theta^2_{\m} (\tau,z) \prod_{\substack{\n \in \pi(V_1) \\ \n \neq \m}} \theta^2_\n + O(q^2).
 \end{align}
Expanding the square root then easily yields
\begin{align}
 P(V_1)^{\frac12} =  \prod_{\m \in \pi(V_1)}  \theta_\m - 2 q \sum_{\m \in \pi(V_1)} \frac{\theta^2_{\m} (\tau,z)}{\theta^2_{\m}(\tau,0)} \prod_{\n \in \pi(V_1) } \theta_{\n} + O(q^2).
 \end{align}
Finally, we use the heat equation for the theta functions, where $\delta_{ij}$ is the Kronecker delta,
\begin{align}
 \pdd{\theta_\m}{z_i}{z_j} = 2 \pi i (1+\delta_{ij}) \pd{\theta_\m}{\tau_{ij}}
\end{align}
and the fact that $\theta_{\m}(z)$ is an even function of $z$ whenever $\m$ is an even characteristic, to obtain
\begin{align}
 \left. \pdd{P(V_1)^{\frac12}}{z_i}{z_j} \right|_{z=0} = - 8 \pi i (1 + \delta_{ij}) q \br{ \sum_{\m \in \pi(V_1)} \pd{\theta_\m}{\tau_{ij}} \prod_{\n \neq \m} \theta_{\n} } + O(q^2).
\end{align}
Note that $P(V_1)$ is an even function of $z$ and thus the odd partial derivatives vanish (up to $O(q^2)$).

 \item Next, let $V_2 \in \charspaces{5}{5}$ contain both characteristics of the form $\m=\bmat{ 0 & \epsilon \\ 0 & \delta }$ and of the form $\m=\bmat{ 1 & \epsilon \\ 0 & \delta }$, but none with $\dfm = 1$.

If there is at least one element $\m \in V_2$ such that $\efm=1$ it is easy to see that for exactly half of the elements $\n \in V_2$ we will have $\efn=1$ while for the other half we will have  $\efn =0$.
Therefore, using \eqref{eq:FJdelta0} and \eqref{eq:FJdelta1} to expand all theta constants, we have
 \begin{align}
  P(V_2) = 2^{16} q^2 \prod_{\substack{\m \in V_2 \\ \efm = 0}} \theta_{\prm}(\tau,0) \prod_{\substack{\n \in V_2 \\ \efn = 1}} \theta_{\prn}(\tau,\frac{z}2) + O(q^3)
 \end{align}

Similar to case 1) above, the $\prm$ are pairwise equal and the corresponding pairs of $\m$ differ only in the component $\efm$. Thus, we end up with
\begin{align}
 P(V_2)^{\frac12} = 2^8 q \sqrt{\prod_{\m \in \pi(V_2)} \theta_{\m} (\tau,0)\, \theta_{\m}(\tau,\frac{z}2)} + O(q^2).
\end{align}
Also, 
recalling that all $\m \in \pi(V_2)$ are even and applying the theta heat equation we find
\begin{align}
 \left. \pdd{P(V_2)^{\frac12}}{z_i}{z_j} \right|_{z=0}
 &= 2^5 q \sum_{\m \in \pi(V_2)} \pdd{\theta_{\m}}{z_i}{z_j} \prod_{\n \neq \m} \theta_{\n} + O(q^2) \\
 &= 64 \pi i (1 + \delta_{ij}) q \sum_{\m \in \pi(V_2)} \pd{\theta_{\m}}{\tau_{ij}} \prod_{\n \neq \m} \theta_{\n} + O(q^2).
\end{align}
Note that $P(V_2)^{\frac12}$ is an even function of $z$ and thus the odd partial derivatives vanish (up to $O(q^2)$).

 \item \label{thirdkind} Last, we consider subspaces containing, in addition to characteristics contained in subspaces from case 2) above, characteristics of the form $\m= \bmat{  0 & \epsilon \\ 1 & \delta }$. These do not have the simple pairings observed above, but we can still expand the theta constants and obtain the similar expression below, but it cannot be simplified as easily. This, however, will turn out not to be necessary for our purposes. The 16 factors of $e^{\pi i \dfm}$ together yield 1, and we end up with
\begin{align}
P(V_3)^{\frac12} = 2^{8} q \sqrt{\prod_{\substack{\m \in V_3 \\ \efm = 0}} \theta_{\prm}(\tau,0) \prod_{\substack{\n \in V_3 \\ \efn = 1}} \theta_{\prn}(\tau,\frac{z}2)} + O(q^2).
\end{align}
For any genus $g$ there will be at least $2^{g-2}$ odd characteristics in $\pi(V_3)$ when $V_3$ is of this type. Therefore, we have
\begin{align}
 \left. \pd{P(V_3)^{\frac12}}{z_i} \right|_{z=0} =
 \left. \pdd{P(V_3)^{\frac12}}{z_i}{z_j} \right|_{z=0} =
 \left. \frac{\partial^3 P(V_3)^{\frac12}}{\partial z_i \partial z_j \partial z_k} \right|_{z=0} = 0
\end{align}
at least up to $O(q^2)$.

\end{enumerate}

\subsubsection{The expression for $G_5^{(5)}$}
Let $\mathcal{V}_k$ be the subset of $\charspaces{5}{5}$ containing all subspaces from case $k$) above.
 Note that $\pi(V)$, for $V \not\in \mathcal{V}_3$, is a totally isotropic element of $\charspaces{4}{4}$, and in fact $\pi(\mathcal{V}_1)=\pi(\mathcal{V}_2)$ is the set of all totally isotropic elements of $\charspaces{4}{4}$, so $G_5^{(5)} = G_4^{(4)} + O(q)$.

Now, combining the results from the previous subsection,
\begin{align}
G_5^{(5)} &= \sum_{V \in \charspaces{5}{5}} P(V)^{\frac12} \nonumber \\
&= G_4^{(4)} + 2^8 \, q \left( \vphantom{\sqrt{\prod_{\substack{\m \in V_3 \\ \efm = 0}} \theta_{\prm} \prod_{\substack{\n \in V_3 \\ \efn = 1}}}} \sum_{V \in \charspaces{4}{4}}  \sqrt{\prod_{\m \in V} \theta_{\m} \, \theta_{\m}(\tau,\frac{z}2)} - 2^{-7} \sum_{\m \in V} \frac{\theta^2_{\m} (\tau,z)}{\theta^2_{\m}} \prod_{\n \in V } \theta_{\n}  \right. \nonumber \\
 & \qquad \left. + \sum_{V_3 \in \mathcal{V}_3} \sqrt{\prod_{\substack{\m \in V_3 \\ \efm = 0}} \theta_{\prm} \prod_{\substack{\n \in V_3 \\ \efn = 1}} \theta_{\prn}(\tau,\frac{z}2)} \right) + O(q^2).
  \end{align}
Also, this gives us
\begin{align}
 \left. \pdd{G_5^{(5)}}{z_i}{z_j}\right|_{z=0} &= 56 \pi i (1 + \delta_{ij}) q \sum_{V \in \charspaces{4}{4}} \sum_{\m \in V} \pd{\theta_\m}{\tau_{ij}} \prod_{\n \neq \m} \theta_\n + O(q^2) \\
 &= 56 \pi i (1 + \delta_{ij}) q \pd{G_4^{(4)}}{\tau_{ij}} + O(q^2). \label{eq:g55zij}
\end{align}
And finally, as the contribution from all $V_3 \in \mathcal{V}_3$ will vanish in $z=0$ because $\pi(V_3)$ contains odd characteristics, we can see that
\begin{align}
 \left. G_5^{(5)} \right|_{z=0} =  \br{1 + 224\,q } G_4^{(4)}  + O(q^2) \label{eq:G55q8z0}.
\end{align}
Note that, because the first terms from the expansion of $G_1^{(1)}(\lambda)$ are $1 + 224\,q$, this is consistent with the factorization property for $G_g^{(g)}$.

\subsection{The expansion of $\vartheta_5^{(5)}$}
We will now do the same for $\vartheta_5^{(5)}$ as done above for $G_5^{(5)}$, that is, take the Fourier-Jacobi expansion and calculate the $z_i,z_j$ derivatives of the first terms.

Note that as $\vartheta_5(\tau)^{(g)} := \sum_{p_1,\dots,p_g \in \Lambda_5} e^{\pi i (p_k\cdot p_l)\tau_{kl}}$, we can write
\begin{align}
\vartheta_5 \pmat{ \lambda & z^t \\ z & \tau } =  \sum_{p_1,\dots ,p_5 \in \Lambda_5} e^{\pi i p_1 \cdot p_1 \lambda} e^{2 \pi i \sum_i p_1 p_i z_i} e^{\pi i \sum_{i,j>1}^5 p_i p_j \tau_{ij}}.
\end{align}
The first term in the $q$-expansion is easy to obtain, and we will obtain the $q$-linear term as in \cite{matone2010getting} by writing
\begin{align}
F^{(g)}(\tau,z) := \sum_{p_1,\dots ,p_g \in \br{D_8 \oplus D_8}^+} e^{\pi i \sum_{i,j=1}^g p_i p_j \tau_{ij}} \sum_{\tilde{p} \cdot \tilde{p} = 2} e^{2 \pi i  \sum_{i=1}^g \tilde{p} p_i z_i}
\end{align}
Clearly, the norm 2 vectors are $(\ldots,\pm 1,\ldots,\pm 1,\ldots,0^8)$ and $(0^8,\ldots,\pm 1,\ldots,\pm 1,\ldots)$, where $\ldots$ denotes a possibly empty sequence of zeroes. There are $2\cdot 4 \cdot \binom82 = 224$ of those.

Now the first terms of the series in $q$ will be:
\begin{align}
\vartheta_5^{(5)} \pmat{ \lambda & z^t \\ z & \tau } = \vartheta_5^{(4)}(\tau) + q F^{(4)} (\tau,z) + O(q^2) \label{eq:vartheta5inq}.
\end{align}

Now we will express the $z_i z_j$-derivatives of $F^{(4)}$, the $q$-linear term from \eqref{eq:vartheta5inq}, as done above for $G_5^{(5)}$. Because the norm 2 vectors are the same as those from $D_8$, we can use the fact that
\begin{align}
 \sum_{\tilde{p} \in \br{D_8 \oplus D_8}^+: \tilde{p} \cdot \tilde{p}=2} (p_i \cdot \tilde{p}) (p_j \cdot \tilde{p}) = 28\, p_i \cdot p_j,
\end{align}
which is mentioned and used in \cite{matone2010getting}. We then obtain
\begin{align}
 \left. \pdd{F^{(4)}}{z_i}{z_j} \right|_{z=0} &= \sum_{p_1,\dots ,p_4 \in \Lambda_5} e^{\pi i \sum_{i,j=1}^g p_i p_j \tau_{ij}} \sum_{\tilde{p} \cdot \tilde{p} = 2} (2 \pi i)^2 (\tilde{p} p_i) (\tilde{p} p_j) \\
 &=  56 \pi i(1 + \delta_{ij})  \left. \pd{F^{(4)} }{\tau_{ij}} \right|_{z=0} =  56 \pi i(1 + \delta_{ij})   \pd{\vartheta_5^{(4)}}{\tau_{ij}} \label{eq:vartheta5q8zij}.
\end{align}


\subsection{The final expression}
Let now, for brevity, $f^{(g)},f_0^{(g)}$ and $f_1^{(g)}$ be defined by
\begin{align}
 f^{(g)} := \vartheta_5^{(g)} - G_{g}^{(g)} \\
 f^{(g)} = f_0^{(g)} + q f_1^{(g)} + O(q^2).
\end{align}
We now develop $f^{(5)}$ as a function of $s$. Applying formula \eqref{eq:sseries} to $f^{(5)}$ and noting that $f_0^{(5)} = f^{(4)}$, we have
\begin{align}
 f^{(5)}(\tau_s) = f^{(4)}(\tau_0) + s\br{ f_1^{(5)}(\tau_0,z) + \sum_{i \leq j} \pd{f^{(4)}}{\tau_{ij}} \sigma_{ij}(p,p')} + O(s^2).
\end{align}

Now, we expand this using \eqref{eq:useries}, letting $u := x(p) - x(p')$ for a local chart $x$. For brevity we write
\begin{align}
S_{ij} := \frac{u^2}{4}  \pd{\omega_i(p)}{x} \pd{\omega_j(p)}{x} + \frac{u^3}2 \pdd{\omega_i(p)}{x}{} \pd{\omega_j(p)}{x}.
\end{align}
Remember that $\sigma_{ij}(p,q) = S_{ij} + O(u^4)$. Then,
\begin{align}
 f^{(5)}(\tau_s) = f^{(4)}(\tau_0) + s \sum_{i \leq j} \br{ u^2 \pdd{f_1^{(5)}}{z_i}{z_j} \omega_i(p) \omega_j(p) + \pd{f^{(4)}}{\tau_{ij}} S_{ij} +  O(u^4)} + O(s^2).
\end{align}
By \eqref{eq:g55zij} and \eqref{eq:vartheta5q8zij} we know that $\pdd{f_1^{(5)}}{z_i}{z_j} = 56 \pi i(1 + \delta_{ij}) \pd{f^{(4)}}{\tau_{ij}}$. This leaves us with
\begin{align}
 f^{(5)}(\tau_s) = f^{(4)}(\tau_0) + s \sum_{i \leq j}  \pd{f^{(4)}}{\tau_{ij}} \br{ 56 \pi i(1 + \delta_{ij}) u^2 \omega_i(p) \omega_j(p) + S_{ij} +  O(u^4)} + O(s^2).
\end{align}
Now, let $J^{(g)} := \vartheta_6^{(g)} - \vartheta_7^{(g)}$. Because $f^{(4)} = \frac37 J^{(4)}$, from \cite{latticevsriemann}, we can rewrite the above as follows:
\begin{align}
  f^{(5)}(\tau_s) = \frac37 J^{(4)}(\tau_0) + \frac{3s}7 \sum_{i \leq j}  \pd{J^{(4)}}{\tau_{ij}} \br{ 56 \pi i(1 + \delta_{ij}) u^2 \omega_i(p) \omega_j(p) + S_{ij} +  O(u^4)} + O(s^2).
\end{align}

In \cite[p.~16-17]{grushevsky2011superstring} Grushevsky and Salvati Manni obtain a similar expression for the degeneration of $J^{(5)}$, differing only in the numerical coefficients. They show that the $\omega_i(p)\omega_j(p)$ term vanishes and that $\sum_{i \leq j}  \pd{J^{(4)}}{\tau_{ij}} S_{ij}$ cannot vanish everywhere due to the fact that $J^{(4)}$ is the Schottky form. We refer to \cite{grushevsky2011superstring} for details. This shows that $f^{(5)}(\tau_s)$ does not vanish everywhere. Thus, the above leads to the conclusion
\begin{align}
 \vartheta_5^{(5)} \not\equiv G_5^{(5)}
\end{align}
when restricted to $\mathcal{J}_5$, as promised. \qed

The factorization properties of $G_g^{(g)}$ and $\vartheta_5^{(g)}$ now imply that this holds for higher genera as well - assuming $G_g^{(g)}$ is well-defined for $g \geq 6$, the above implies that $f^{(g)}$ does not vanish identically on $\mathcal{J}_g$, for all $g \geq 5$.

\section{The trace of $f^{(5)}$} \label{sec:trace}
Here we will look at the trace of $f^{(5)}$, defined as $\sum_\m f^{(5)}[\m]$, because it occurs in the cosmological constant and is thus of interest for the genus 5 measure.

The definition of $f^{(5)}[\m]$ is as follows: for any modular form $f$ and for $\gamma_\m = \pmat{ A & B \\ C & D}$ such that $\bmat{ \diag(A^T C) \\ \diag(B^T D) }  =  \m$, we have $f[\m] := (f|_8 \, \gamma_\m)$. When $f$ is a modular form with respect to $\Gamma(1,2)$, $f[\m]$ does not depend on the particular choice of $\gamma_\m$.

In \cite{grushevsky2011superstring} Grushevsky and Salvati Manni calculate the traces of the forms $G_p^{(g)}$. They use a different notation: their $S_i$ equals $2^{-i} \sum_\m G_i [\m]$. Their result shows that all $\sum_m G_p^{(g)}$ can be recursively expressed through $\sum_m G_0^{(g)} $ and $\sum_m G_1^{(g)}$. Note that this formula is only valid for the $G_p^{(g)}$ with $p \leq g$, because the others vanish identically, and for $1 \leq n \leq 4$.
\begin{align}
2  \sum_m G_{n+1}^{(g)}[m]=\frac{ 2^{2 (g-n+1)}-1}{\left(2^n-1\right) \left(2^{n+1}-1\right)} \sum_m G_{n-1}^{(g)}[m] -\frac{3}{2^{n+1}-1}  \sum_m G_{n}^{(g)}[m]
\end{align}
Because $G_0^{(5)}[\m] = \theta_\m^{16}$ and $G_1^{(5)}[\m] = \theta_\m^8 \sum_{\n \neq 0} \theta_{\m + \n}^8$, we see that $\sum_\m G_0^{(5)}[\m] = \sum_\m \theta_\m^{16} = \vartheta_7 $, and $\sum_\m G_1^{(5)}[\m] = \br{ \sum_\m \theta_\m^8 }^2 - \sum_\m \theta_\m^{16} = \vartheta_6 - \vartheta_7$. Therefore, we can easily obtain
\begin{align}
 \sum_\m G_5^{(5)}[\m] = \frac{32}{217} \br{950 \, \vartheta_6^{(5)} - 733 \, \vartheta_7^{(5)}}\\
 \sum_\m G_4^{(4)}[\m] = -\frac{16}{7} \br{22 \, \vartheta_6^{(4)} - 29 \, \vartheta_7^{(4)}}.
\end{align}
From \cite[p.~28]{matone2010getting} we learn that
\begin{align}
 \sum_{\m} \vartheta_5^{(g)} [\m] = 2^{g-1} \br{ \vartheta_6^{(5)} + \vartheta_7^{(5)} }.
\end{align}
Combining the above facts, we obtain the following expressions for the genus 4 and genus 5 trace of $f^{(g)}$:
\begin{align}
  \sum_\m f^{(4)}[\m] = - \frac{2^3 \cdot 3 \cdot 17}{7} J^{(4)} \label{eq:genus4trace}\\
  \sum_\m f^{(5)}[\m] =  \frac{2^4 \cdot 3^2 \cdot 11 \cdot 17}{7 \cdot 31} J^{(5)}. \label{eq:genus5trace}
\end{align}

This implies that although $f^{(5)}$ is a cusp form with respect to $\Gamma(1,2)_5$, $\sum_m f^{(5)}[m]$ is a cusp form with respect to all of $\Gamma_5$. Note that because on $\overline{\mathcal{M}_5}$ there exists a unique divisor of slope 8 \cite{harris1990slopes}, any cusp form of weight 8 with respect to all of $\Gamma_5$ will be proportional to $J^{(5)}$, so \eqref{eq:genus5trace} is not surprising. Of course, \eqref{eq:genus4trace} is not surprising either, as there is only one form vanishing on all of $J^{(4)}$; we are just interested in the specific coefficients of \eqref{eq:genus4trace} and \eqref{eq:genus5trace} for the purpose of the next section.

In genus $g$ there are $2^{g-1}(2^g + 1)$ even characteristics. Because $J^{(g)}$ is a modular form with respect to the entire modular group $\Gamma_g$, its trace is simply the number of even characteristics times $J^{(g)}$. Note that $\frac{\sum_\m f^{(5)}[\m]}{\sum_\m f^{4}[\m]} \neq \frac{\sum_\m J^{5}[\m]}{\sum_\m J^{4}[\m]}$. This fact will be used in Section \ref{sec:twopointfunction} to obtain both a vanishing cosmological constant in genus 5 \emph{and} a vanishing two-point function in genus 4; in \cite{matone2010getting} it was shown that it is impossible to do this using only the OPSMY forms while conforming to the other requirements for the measure.

\emph{Remark}. Note that if $f^{(5)}$ were to vanish on $\mathcal{J}_5$, this would imply that the trace would vanish as well. Since $J^{(5)}$ is not everywhere zero on $\mathcal{J}_5$, see \cite{grushevsky2011superstring}, this gives a second proof that $f^{(5)}$ does not vanish identically there.

\section{The difference between $f^{(5)}$ and $J^{(5)}$} \label{sec:difference}
Now that we know that $f^{(5)}$ does not vanish everywhere on $\mathcal{J}_5$, a natural question which arises is whether this form is linearly independent from the already known modular forms with respect to $\Gamma(1,2)$ on $\mathcal{J}_5$. By the factorization property for both the Grushevsky and OPSMY basis, we can eliminate all but one candidate. Because (from \cite{latticevsriemann}) $f^{(4)} = \frac37 J^{(4)}$, we see that
\begin{align}
 f^{(5)} \pmat{ \lambda & 0 \\ 0 & \tilde{\tau}} = \vartheta_5^{(1)} f^{(4)}= \frac37 \vartheta_5^{(1)}  J^{(4)}. \label{eq:f5factorization}
\end{align}

All lattice theta series have the simple factorization property $\vartheta_p^{(g)}(\tau_1 \oplus \tau_2) = \vartheta_p^{(k)}(\tau_1) \vartheta_p^{(g-k)} (\tau_2)$ and all other known modular forms with respect to $\Gamma(1,2)$ can be expressed through them \cite{latticevsriemann}. The only linear combination of these forms yielding the same as \eqref{eq:f5factorization} is equal to $J^{(5)}$.


We will prove by a simple argument that $f^{(5)}$ and $J^{(5)}$ cannot coincide on the Jacobian locus $\mathcal{J}_5$. Recall that
\begin{align}
 \sum_\m f^{(5)}[\m] = \frac{3 \cdot 17}{7 \cdot 31}\sum_\m J^{(5)}[\m]. \label{eq:tracesf5andj5}
\end{align}
Looking at the degeneration found in section \ref{sec:degeneration},
\begin{align}
f^{(5)} = f^{(4)} +
 \frac37 s \sum_{i \leq j} \pd{J^{(4)}}{\tau_{ij}} \left( 56 u^2 (1+\delta_{ij})\omega_i(q) \omega_j(q) +  u^2 \frac14 \pd{\omega_i}{x}(q) \pd{\omega_j}{x}(q) \right. \nonumber \\
 \left. + \frac12 u^3 \pdd{\omega_i(q)}{x}{} \pd{\omega_j(q)}{x} + O(u^4) \right) + O(s^2), \label{eq:tudegen1}
\end{align}
we can compare it with the very similar expression found in \cite{grushevsky2011superstring} for the first terms in $u$ in the $s$-linear term when taking the same degeneration for $J^{(5)}$,
\begin{align}
 J^{(5)} = J^{(4)} +
 s \sum_{i \leq j} \pd{J^{(4)}}{\tau_{ij}} \left( 30 u^2(1+\delta_{ij})\omega_i(q) \omega_j(q) +  u^2 \frac14 \pd{\omega_i}{x}(q) \pd{\omega_j}{x}(q)  \right. \nonumber \\
 + \left.  \frac12 u^3 \pdd{\omega_i(q)}{x}{} \pd{\omega_j(q)}{x} + O(u^4) \right) + O(s^2). \label{eq:tudegen2}
\end{align}
Because, from \cite{grushevsky2011superstring},
\begin{align}
 \sum_{i \leq j} \pd{J^{(4)}}{\tau_{ij}}u^2(1+\delta_{ij})\omega_i(q) \omega_j(q) = 0 \\
 f^{(4)} = \frac37 J^{(4)},
\end{align}
we conclude that the only linear combination of $f^{(5)}$ and $J^{(5)}$ that vanishes at first order along the boundary is $f^{(5)} - \frac37 J^{(5)}$, a modular form with respect to $\Gamma(1,2)_5$. But we know from equation \eqref{eq:tracesf5andj5} that $f^{(5)} \neq \frac37 J^{(5)}$ on the Jacobian locus. Therefore, $f^{(5)}$ cannot be a multiple of $J^{(5)}$ everywhere on $\mathcal{J}_5$.


\section{The two-point function in genus 4} \label{sec:twopointfunction}
Matone and Volpato show in \cite{matone2010getting} that it is not possible to make a genus 5 measure from the OPSMY forms that satisfies all requirements, assuming their method of obtaining the genus 4 two-point function is correct. To be precise, the degeneration to genus 4 yields a not identically vanishing two-point function if the genus 5 cosmological constant is made to vanish, i.e. requirement c) from the introduction is not satisfied. Therefore, one may ask whether by combining these forms with $G_5^{(5)}$ one can construct a measure that does satisfy these properties. The answer is yes.

In order to obtain the genus 4 two-point function from the genus 5 measure, we follow the procedure set by Matone and Volpato. Therefore, all results in this section depend on their procedure being correct. A discussion of its validity is, however, beyond the scope of the present paper. For details, we refer to the original paper \cite{matone2010getting}. That is, consider
\begin{equation}
X_{NS}[(\epsilon,\delta)] := \frac12\br{ \txi^{(g+1)} \bmat{0 & \epsilon \\ 0 & \delta}  + \txi^{(g+1)} \bmat{0 & \epsilon \\ 1 & \delta}}
\end{equation}
 and contract one handle from a family of curves, where then the term linear in the perturbation parameter will be the two-point function. As the argument from \cite{matone2010getting} is quite detailed, we will just look at what happens with the terms $c_J J^{(5)}+ c_f f^{(5)}$ which we would like to add to the measure, instead of $-B_5  J^{(5)}$ as originally proposed, where $B_5$ is the coefficient of $ J^{(5)} $ in the cosmological constant from the 'plain' OPSMY ansatz. From the degeneration in the limit $s \rightarrow 0$, we obtain a surface with two nodes $a$ and $b$. Now, let $\nu_*^2(c) = \partial_i \theta_*(0) \omega_i(c)$ for an odd theta
characteristic $*$ and define
\begin{align}
 E(a,b) := \frac{\theta_*(A_{ab})}{\nu_* (a) \nu_* (b)}
\end{align}
which is the prime form, see \cite{fay}. Let $A_2[\m](a,b)$ be the two-point function. We will have up to a factor independent of $e$, in some choice of local coordinates,
\begin{align}
 X_{NS}[\m] = s E(a,b)^2 A_2[\m](a,b)+ O(s^2),
\end{align}
from \cite{matone2010getting}. For the OPSMY part of the ansatz we will stick to the notation from Matone and Volpato, that is, we will write $\Theta_k$ for the lattice theta series, with a different numbering of lattices for $k \leq 5$, so that it is easier to compare the formulae. Here we present a translation diagram:
\begin{center}
\begin{tabular}{|c|c|c|}
\hline
\cite{matone2010getting} notation&Lattice&Our notation\\
\hline
$\Theta_0$&$\br{D_8\oplus D_8}^+$&$\vartheta_5$\\
\hline
$\Theta_1$&$\Z\oplus A_{15}^+$&$\vartheta_4$\\
\hline
$\Theta_2$&$\Z^2\oplus \br{E_7\oplus E_7}^+$&$\vartheta_3$\\
\hline
$\Theta_3$&$\Z^4\oplus D_{12}^+$&$\vartheta_2$\\
\hline
$\Theta_4$&$\Z^8\oplus E_{8}$&$\vartheta_1$\\
\hline
$\Theta_5$&$\Z^{16}$&$\vartheta_0$\\
\hline
$\Theta_6$&$E_8 \oplus E_8$&$\vartheta_6$\\
\hline
$\Theta_7$&$D_{16}^+$&$\vartheta_7$\\
\hline
\end{tabular}
\end{center}
Let $N_k$ be the number of norm two vectors in the lattice corresponding to $\Theta_k$. Let $c_k^g$ be the coefficient of $\Theta_k$ in the OPSMY ansatz for genus $g$, where the same normalization as in \cite{grushevsky} is used ($c_k^g$ is $2^{4g}$ times the coefficients from \cite{oura2010modular}) for easier comparison.

We have, for the OPSMY ansatz, from \cite{matone2010getting},
\begin{align}
 X_{NS}[\m](s,\Omega,z=0) = \sum_{k=0}^7 c_k^{5} \br{1 + N_k s + O(s^2)} \Theta_k^{(4)}[\m](\Omega).
\end{align}
We will write
\begin{align}
 X_{NS}[\m](s,\tau,z) = T_0[\m](\tau,z) + s \, T_1[\m](\tau,z)+ O(s^2).
\end{align}
Note that $E_8 \oplus E_8$ and $D_{16}^+$ contain $480$ norm 2 vectors and $\br{D_8 \oplus D_8}^+$ contains 224 of them. Also, the $s$-linear term from $G_5^{(5)}$, formula \eqref{eq:G55q8z0}, equals $244\, G_4^{(4)}$ in $z=0$. Therefore, we have
\begin{align}
T_0[\m](\tau,0) &= \sum_{k=0}^5 c_k^{5}\Theta_k^{(4)}[\m] + cJ^{(4)}+ c_f f^{(4)}  = \br{c_J - \frac{2^5 \cdot 3}7} J^{(4)} + c_f  f^{(4)} \\
 T_1[\m](\tau,0) &= 128 \Xi_{OPSMY}^{(4)}[\m](\tau) +  \br{480 c_J - \frac{720 \cdot 2^5 \cdot 3}7} J^{(4)} + 224 c_f  f^{(4)}
\end{align}
As $s \rightarrow 0$, we get
\begin{align}
 X_{NS}[\m] &= s \sum_{i,j}^4 2 \pi i E(a,b)^2 \omega_i(a) \omega_j(b)(1 +  \delta_{ij})\br{  \br{c_J - \frac{2^5 \cdot 3}7}\pd{ J^{(4)}}{\tau_{ij}} + c_f \pd{f^{(4)}}{\tau_{ij}} } \nonumber \\
 & \qquad + s T_1^{(4)}[\m](\tau,A_{ab})  + O(s^2).
\end{align}
Calculating $T_1[\m](\tau,A_{ab})$ from $T_1[\m](\tau,0)$ can be done using the fact that $T_1[\m]$ is a section of $|2\Theta|$, because of the modular properties of $X_{NS}$. Here $\Theta$ is the divisor of $\theta_0 (z)$. Matone and Volpato prove that from that fact it follows that

\begin{align}
 T_1[\m](\tau,A_{ab}) = E(a,b)^2 \br{T_1[\m](\tau,0) \omega(a,b) + \frac12 \sum_{i,j}^4 \partial_i \partial_j T_1[\m](\tau,0) \omega_i(a) \omega_j(b) }.
\end{align}
From \cite{latticevsriemann}, we have $f^{(4)} = \frac37 J^{(4)}$ which is the Schottky form and vanishes on $ \mathcal{J}_4$. Thus we have $T_1[\m](\tau,0) = 128 \, \Xi_{OPSMY}^{(4)}$ on the Jacobian locus. Then, we get
\begin{align}
 A_2[\m](a,b) &= 128 \, \Xi^{(4)}[\m](\tau)\omega(a,b) +  \sum_{i,j}^4 2 \pi i(1+ \delta_{ij}) \omega_i(a) \omega_j(b)\, \cdot \nonumber \\
   & \qquad \cdot \br{ \br{ c_J - \frac{2^5 \cdot 3}7}\pd{ J^{(4)}}{\tau_{ij}} + c_f \pd{f^{(4)}}{\tau_{ij}}  + \frac12 \partial_i \partial_j T_1^{(4)}[e](\tau,0)} .
\end{align}
Denoting by $f_1^{(5)}$ the $s$-linear term from the $s$-expansion of $f^{(5)}$, and using the functions
\begin{align}
F_k^{(g)}(\tau,z) := \sum_{p_1,\dots ,p_g \in \Lambda_k} e^{\pi i \sum_{i,j=1}^g p_i p_j \tau_{ij}} \sum_{\tilde{p} \cdot \tilde{p} = 2} e^{2 \pi i  \sum_{i=1}^g \tilde{p} p_i z_i}
\end{align}
we end up with the modified formula
\begin{align}
 \partial_i \partial_j T_1^{(4)}[\m](\tau,0) = \partial_i \partial_j  \br{\sum_{k=0}^5 c_k^{5} F_k^{(4)}[\m](\tau,0) + c_J \br{ F_6^{(4)} - F_7^{(4)}} + c_f f_1^{(5)} }.
\end{align}
Here, Matone and Volpato introduce the coefficients $s_k^g$ and $t_k^g$, defined by the following formula:
\begin{align}
  \partial_i \partial_j c_k^{g+1} F_k^{(g)}[\m](\tau,0) = 2 \pi i(1 + \delta_{ij}) \partial_i \partial_j s_k^g \Theta_k^{(g)}[\m] - t_k^g \Theta_k^{(g)} \partial_i \partial_j \log \theta[\m](\tau,0).
\end{align}
Continuing the process from \cite{matone2010getting}, and noting that $f_1^{(5)}$ has the property that $\pdd{f_1^{(5)}}{z_i}{z_j} = 28( 2 \pi i)(1 + \delta_{ij})  \pd{f^{(4)}}{\tau_{ij}}$ (see formulae \eqref{eq:g55zij} and \eqref{eq:vartheta5q8zij}), we then get
\begin{align}
 \partial_i \partial_j T_1^{(4)}[\m](\tau,0) = 2 \pi i(1 + \delta_{ij}) \pd{}{\tau_{ij}} \br{ \sum_{k=0}^5 s_k^4 \Theta_k^{(4)} [e](\tau) + 60 c_J  J^{(4)} + 28 c_f f^{(4)}  } \nonumber \\
 -  \br{ \sum_{k=0}^5 t_k^4 \Theta_k^{(4)} [\m](\tau) } \partial_i \partial_j \log \theta[\m](\tau,0).
\end{align}
And further following the calculations from \cite{matone2010getting} the first term in big brackets can be written as
\begin{align}
\begin{split} \sum_{k=0}^5 s_k^4 \Theta_k^{(4)} [\m](\tau) + 60 c_J J^{(4)} + 28 c_f f^{(4)} &=  32 \Xi^{(4)}[\m](\tau) 
 \\&+ \br{60 c_J + \frac{3 \cdot 28}7 c_f - \frac{152 \cdot 2^5 \cdot 3}7 }J^{(4)}.
\end{split}
\end{align}
So, having carried the modified $\txi$ through the degeneration, we end up with a slightly different two-point function,
\begin{align}
 \begin{split}
 A_2[\m](a,b) &= 128 \Xi^{(4)}[\m](\tau)\omega(a,b) \\
 & - \sum_{i,j}^4 \omega_i(a) \omega_j(b) \Bigg[128 \Xi^{(4)}[\m](\tau) \partial_i \partial_j \log \theta[\m](\tau,0) 
 \\ &\qquad - 2 \pi i(1 + \delta_{ij}) \pd{}{\tau_{ij}} \Bigg\{ 16 \Xi^{(4)}[e](\tau)  \\
&\qquad \qquad + \br{(30+1) c_J + (6+1) c_f - \frac{(76+1) \cdot 2^5 \cdot 3}7 } J^{(4)} 
 \Bigg\} \Bigg]
 \end{split}
\end{align}
The last step of the procedure from \cite{matone2010getting} is to sum over even characteristics. This procedure yields, finally, on $\mathcal{J}_4$,
\begin{align}
 \sum_\m A_2[\m](a,b) &= 2^3(2^4 + 1) \, C_4 \sum_{i,j}^4 \omega_i(a) \omega_j(b) 2 \pi i(1 + \delta_{ij}) \pd{J^{(4)}}{\tau_{ij}} \\
C_4 :&= \br{16 B_4 - 8 D_4 - 77\frac{2^5 \cdot 3}7 + 31 c_J + 7 c_f}.
\end{align}
So, to make $\sum_\m A_2[\m](a,b)$ vanish, we would need
\begin{align}
 31 c_J + 7 c_f = 77\frac{2^5 \cdot 3}7 + 8 \frac{2^7 \cdot 3}{7 \cdot 17} - \frac{2^6 \cdot 3^3 \cdot 5 \cdot 11}{7 \cdot 17}. \label{eq:genus5lineq1}
\end{align}

The genus 5 cosmological constant from the 'plain' OPSMY ansatz, that is, without the $-B_5 J^{(5)}$ part, equals (again, see \cite{matone2010getting}),
\begin{align}
  \sum_\m \sum_{k=0}^5 c_k^5 \Theta_k[\m] = - 2^4(2^5 + 1) \frac{2^5 \cdot 17}{7 \cdot 11} J^{(5)}.
\end{align}
From Section \ref{sec:trace}, we have for the trace of $f^{(5)}$:
\begin{align}
 \sum_\m f^{(5)}[\m] = \frac{2^4 \cdot 3^2 \cdot 11 \cdot 17}{7 \cdot 31} J^{(5)}.
\end{align}
Because $E_8 \oplus E_8$ and $D_{16}^+$ are even lattices, they are invariant under modular transformations and therefore
\begin{align}
 \sum_\m J^{(5)}[\m] = 2^4(2^5 + 1)J^{(5)}.
\end{align}
Thus, to make the genus 5 cosmological constant vanish we would need
\begin{align}
 2^4(2^5 + 1)c_J + \frac{2^4 \cdot 3^2 \cdot 11 \cdot 17}{7 \cdot 31}c_f = 2^4(2^5 + 1) \frac{2^5 \cdot 17}{7 \cdot 11}. \label{eq:genus5lineq2}
\end{align}
Combining the above linear equations \eqref{eq:genus5lineq1} and \eqref{eq:genus5lineq2}, we find the solution
\begin{align}
 c_J = -\frac{222647008}{217}, c_f = \frac{77245568}{17}.
\end{align}
 Hence we present our main formula:
\begin{align}
 \boxed{ \txi := \Xi^{(5)}_{OPSMY} - \frac{222647008}{217} J^{(5)}+ \frac{77245568}{17} f^{(5)} }  \label{eq:mainformula}
\end{align}
and the above amounts to proving our main result:
\begin{theorem}
$\txi$ is the unique linear combination of known modular forms of weight 8 that yields both a vanishing genus 5 cosmological constant and a vanishing genus 4 two-point function. \label{maintheorem}
\end{theorem}

\section{The situation in genus 6} \label{sec:genus6}
Here we take a brief look at the current state of the ans\"{a}tze in genus 6 and the possibility of improving it using our findings.

Let $\Xi_\m^{(6)}$ be the Grushevsky ansatz for genus 6\footnote{Note that it is not certain (and there is even no reason to believe) that this is well-defined, as it contains fourth roots of $P(V)$ for $V \in \charspaces{6}{6}$. We know that $G_5^{(5)}$ is well-defined from the proof by Salvati Manni in \cite{salvati2008remarks}.}  (see \cite[Th.22]{grushevsky}). Then, define
\begin{align}
 \txi_\m^{(6)} := \Xi_\m^{(g)} + k_6 f^{(6)} + l_6 J^{(6)}.
\end{align}
For genus 6, the factorization condition gives
\begin{align}
\txi_\m^{(6)}\pmat{\lambda & 0 \\ 0 & \tilde{\tau}} = \Xi_\m^{(5)} \Xi_\m^{(1)} + k_6 \br{\vartheta_5^{(1)} \theta_5^{(5)} - G_1^{(1)} G_5^{(5)} } + l_6 \br{ \vartheta_6^{(1)} \vartheta_6^{(5)} - \vartheta_7^{(1)} \vartheta_7^{(5)}}
\nonumber \\
\mathop{=}^{?} \Xi_\m^{(5)} \Xi_\m^{(1)} +  \Xi_\m^{(1)} \br{k_5 f^{(5)} + l_5 J^{(5)}}
\end{align}
and as $G_1^{(1)} = \vartheta_5^{(1)}$, $\Xi^{(1)} = \frac12 \br{G_0^{(1)} - G_1^{(1)}}$ and $\vartheta_6^{(1)} = \vartheta_7^{(1)} = \sum_\m G_0^{(1)}[\m]$, this implies
\begin{align}
 k_6 G_1^{(1)}[\m] + l_6 \sum_{\n} G_0^{(1)}[\n] = \frac12 (k_5 + l_5) \br{G_0^{(1)}[\m] - G_1^{(1)}[\m]}
\end{align}
and that implies $k_6 = l_6 = k_5 + l_5 = 0$. By theorem \ref{maintheorem} and equation \eqref{eq:mainformula} we have $k_5 + l_5 \neq 0$; so if we want both the genus 4 two-point function and the genus 5 cosmological constant to vanish, this cannot work.

We conclude that to satisfy the factorization constraint in genus 6 while using the proposed modification in genus 5, one needs a new form that degenerates in a way that solves the above problem.

\section{Discussion} \label{sec:discussion}



Our results imply that the space of cusp forms with respect to $\Gamma(1,2)$ on the Jacobian locus $\mathcal{J}_5$ is at least two-dimensional. Adding any such cusp form to all $\Xi[m]$ does not spoil the factorization property. Then there are three conditions on the genus 5 measures left that can put a restriction on this additional cusp form: the vanishing of the cosmological constant in genus 5, of the two-point function in genus 4 and of the three-point function in lower genera. The vanishing of the cosmological constant and of the two-point function allow us to uniquely determine the cusp form that has to be added to $\Xi[m]$, assuming that the space of these cusp forms has dimension no greater than 2. This leaves two open questions: whether this solution is consistent with a vanishing three-point function in lower genera, and whether there are no additional linearly independent cusp forms of this type. A positive answer to the first question would mean that the proposed ansatz satisfies all the requirements up to and including genus 5, while a positive answer to the second question would imply that this ansatz is unique.



\section*{Acknowledgements}
We are very grateful to Gerard van der Geer for his help and numerous comments and to Samuel Grushevsky for vital remarks. We would like to extend our gratitude to Alexei Morozov for his support and for enabling us to present our work at the EINSTEIN conference at SISSA, and to Laura San Giorgi for her diligent reading. We also thank the anonymous referee for many helpful remarks that substantially improved the paper. 

Our work is much indebted to previous work by M. Matone, R. Volpato, R. Salvati Manni, S. Grushevsky and A. Morozov, among many others.

Our work is partially supported by Ministry of Education and Science of the Russian Federation under contract 14.740.11.0677 (P.DB., A.Sl.), by RFBR grants 12-01-00525 (P.DB.) and 10-02-00509 (A.Sl.), by joint grants 11-01-92612-Royal Society (P.DB., A.Sl.) and 12-02-92108-YaF (A.Sl.), by the Russian President’s Grant of Support for the Scientific Schools NSh-3349.2012.2 (P.DB., A.Sl.) and by NWO grant 613.001.021 (P.DB.).

\bibliographystyle{ieeetr}
\bibliography{citations}

\end{document}